\PassOptionsToPackage{dvips}{graphicx}
\documentclass[preprint, sort&compress]{elsarticle}

\usepackage[english]{babel}
\usepackage{amsbsy}
\usepackage{amsmath}
\usepackage{latexsym}
\usepackage[dvips]{graphicx}
\DeclareMathOperator{\ctg}{ctg}

\begin{document}

\title{Transition radiation from electrons with orbital angular momentum}

\author[tpu]{A.S. Konkov\corref{cor1}}
\ead{anatoliy.konkov@gmail.com}

\author[tpu]{A.P. Potylitsyn\corref{cor2}}
\ead{pap@interact.phtd.tpu.ru}

\author[tpu]{M.S. Polonskaya}
\ead{marPS@mail.ru}

\cortext[cor1]{Corresponding author}
\cortext[cor2]{Principal corresponding author}

\address[tpu]{Tomsk Polytechnic University, Lenin Ave. 30, 634050 Tomsk, Russian Federation}

\begin{abstract}
Several experimental groups have recently obtained the so called vortex electrons (electrons with orbital angular momentum (OAM) of $\ell \leq 100\hbar$) with energies of~$\sim 300$\,keV. The gyromagnetic ratio of such electrons becomes proportional to the OAM value, which leads to the corresponding increase of the electron magnetic moment. In this paper we investigate the transition radiation from the ``charge $+$ magnetic moment'' system using the theory of classical electrodynamics. The circular polarization of optical transition radiation amounts up to $\sim 70$\%, which allows to use this effect for the independent measurement of the electron orbital momentum value.
{\sloppy

}
\end{abstract}


\maketitle

\section{Introduction}
Beams of vortex particles are basically beams with phase singularity, which occurs in the absence of intensity and ambiguity of wave front phase in the center of a beam~\cite{Nye}. Wave front of such particles is of the form of spiral, and a wave phase varies from $0$ to $2\pi$ multiply to $\ell$ when moving on a blade of spiral. In this case $\ell$ characterizes the power of ``vortex'' and is called topological charge or orbital angular momentum (OAM). The ``vortex effect'' is achieved by overlap of wave fronts of separate particles in such a way that in each point of space the local wave directions are vortexed relative to a preferred axis (for example, relative to a mean impulse of wave).  

The existence of vortex photon beams was theoretically predicted in work~\cite{Allen} 20 years ago. The authors considered the transformation of laser beams with amplitude of Laguerre-Gauss distribution to the beams with Hermite-Gauss distribution. They showed that in the result of such distribution the photons get OAM, which can be measured mechanically. The study of vortex electrons started later with fundamental work~\cite{bib:B}. The authors demonstrated the occurrence of OAM in beams of free electrons from the solution of Schr\"{o}dinger equation for paraxial wave packets with a vortex phase and found the connection between OAM and magnetic momentum for such particles.           

The authors of recent experiments~\cite{Uchida, Verbeeck, McMorran} have obtained beams of the so called vortex electrons with orbital angular momentum of up to $\ell \sim 100 \hbar$. Direct measurements of the OAM value carried by electrons can be performed determining the mechanical moment transferred to a target if electrons are totally absorbed in it. However, unlike similar measurements with an OAM photon beam, the absorption of electrons in a target is connected with non-uniform energy deposition along both the beam profile and the target depth, which will introduce noticeable errors during OAM measurements.

The accuracy of optical measurements being several orders higher than that of mechanical measurements, it seems undoubtedly promising to determine the electrons OAM values by their radiation characteristics. The main mechanisms of nonrelativistic electron radiation are bremsstrahlung and transition radiation (TR)~\cite{Potylitsyn}. Both mechanisms have already been studied in details both in theory and experimentally. The authors of paper~\cite{Bliokh} have shown that OAM summarizes with the intrinsic electron magnetic moment, which leads to the increase of resultant magnetic moment that is $\ell$ times greater than Bohr magneton $\mu_B = e\hbar/m_{e}c$. As shown in~\cite{McMorran}, a beam of electrons with OAM that is parallel or antiparallel to a momentum can be formed by spatial selection. Since in this case the orbital momentum ${\boldsymbol \ell} = \ell {\bf e}_p$ (where ${\bf e}_p$ is unit momentum vector) quantization axis is determined for electron beams with OAM (it coincides with averaged electron beam momentum $\bf p$), the same quantization axis can be used for the electron magnetic moment ${\boldsymbol \mu} = g\mu_B{\boldsymbol \ell}$ (where $g$ is gyromagnetic ratio) and one can consider the so called longitudinal magnetic momentum orientation.

It is well known that the bremsstrahlung intensity in the range of photon energies $\hbar \omega < \gamma \hbar \omega_p$, where $\gamma$ is electron Lorentz factor, $\omega_p$ is target material plasma frequency, is significantly suppressed as compared to transition radiation yield. In the case of nonrelativistic electrons $(\gamma \sim 1)$, the process of optical TR is more preferable for the investigation of OAM electron radiation characteristics, since while $\hbar \omega_{opt} < \hbar \omega_p$, the intensity of TR is significantly higher than that of bremsstrahlung. Authors of work~\cite{Bal} observed the optical transition radiation characteristics of electrons with kinetic energy as small as $E_k = 80$\,keV. 
\begin{figure}[h]
\center{\includegraphics[width=0.5\linewidth]{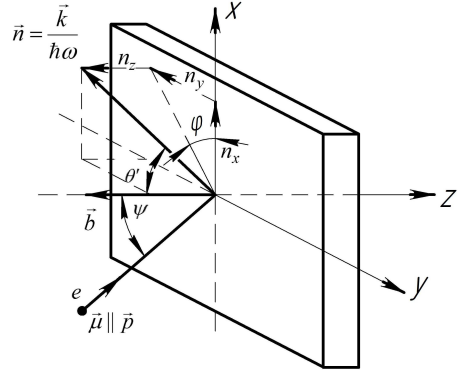}}
\caption{Scheme of TR generation for oblique electron incident on the target}
\label{Fig1}
\end{figure}

In the case of electrons with magnetic moment or OAM, TR will not only be generated by the charge, but by a magnetic moment as well. Considering the spatial parity, the scalar value characterizing the transition radiation process of longitudinal magnetic moment $\boldsymbol \mu$ can be presented as a mixed vector product characterizing a circular polarization of TR:
\begin{equation}
{\boldsymbol \mu} [{\bf b},{\bf k}],\label{eq:1}
\end{equation}
where $\bf k$ is photon momentum, $\bf b$ is a vector perpendicular to the target interface. It is clear that for perpendicular electron beam incident on the target (vacuum-medium interface) the expression~(\ref{eq:1}) is identically zero. However, the circularly polarized radiation component appears with the inclined electron incident with magnetic momentum $\boldsymbol \mu$ on the target of non-coplanar geometry (vector $\bf k$ lies out of the plane $({\boldsymbol \mu}\cdotp {\bf b})$, see fig.1).

\section{Determining the circular polarization degree of OAM electron transition radiation}
For magnetic moment values $|{\boldsymbol \mu}|\gg \mu_B$, the transition radiation of the ``charge $+$ magnetic moment'' system can be considered according to the classical approach with a good accuracy as in paper~\cite{Ginzburg}, where the authors only investigated the perpendicular passage through the interface, when the TR circular polarization was absent and, hence, not considered. In work~\cite{Konkov}, we have obtained the expressions for resulting transition radiated field of the ``charge $+$ magnetic moment'' system directed into the back-hemisphere (around the specular reflection direction -- ``backward TR'') for the target with permittivity $\varepsilon = \varepsilon ' + i\varepsilon ''$. To obtain the results, we used the image method~\cite{Pafomov} that allows obtaining the TR characteristics for any geometry.

For an ideally conducting target~$\left(|\varepsilon ''|\rightarrow \infty \right)$, the fields in the reference frame shown in the figure~1 are as follows:

\begin{align} 
E_x &= C_1 \Bigl(e\beta_{z} n_{x} n_{z} + i\gamma^{-1}\frac{\omega}{c}\mu \Bigl[B_{z}(1 - \beta_{y} n_{y}) n_{y} - \beta_{z} n_{z}^{2}B_{y}\Bigr]\Bigr), \label{eq:0} \\  
E_y &= C_1 \Bigl(e\beta_{z} \Bigl[n_{y} n_{z} - \beta_{y} n_{z}\Bigr] + i\gamma^{-1}\frac{\omega}{c}\mu \Bigl[\beta_{z} n_{z}^{2}B_{x} - B_{z}(1 - \beta_{y} n_{y}) n_{x}\Bigr]\Bigr), \label{eq:0a} \\ 
E_z &= C_1 \Bigl(e\beta_{z} \Bigl[n_{z}^{2} - 1 + \beta_{y} n_{y}\Bigr] + i\gamma^{-1}\frac{\omega}{c}\mu \beta_{z} n_{z} \Bigl[B_{y} n_{x} - B_{x} n_{y}\Bigr]\Bigr). \label{eq:0b}
\end{align}
Here $e$ is electron charge, $\mu$ is magnetic moment,

\begin{gather*}
\begin{split}
{\bf n} = {\bf k}c/\omega &= \{n_x,n_y,n_z\} = \{\sin\theta \cos\phi,\sin\theta \sin\phi,\cos\theta\}, \\
{\boldsymbol \beta} &= \{0,\beta_y,\beta_z\} = \beta \{0,\sin\psi,\cos\psi\}, \\
 B_x &= n_{y}\cos\psi - (\gamma \beta)^{-1}\sin\psi, \\
 B_y &= n_{x}\left(\gamma \sin^{2}\psi - \cos^{2}\psi \right), \\ 
 B_z &= n_{x}(1 + \gamma)\sin\psi \cos\psi, \\
 C_1 &= \frac{\exp\left(i\frac{\omega}{c}R\right)}{\pi cR \left((1 - \beta_{y} n_{y})^{2} - \beta_{z}^{2} n_{z}^{2}\right)}. \\
\end{split}
\end{gather*}
The components of field of TR~(\ref{eq:0}) - (\ref{eq:0b}) were obtained by the method of images according to work~\cite{Konkov} using the known relations for the connection of electric force vector with Hertz potentials, both for electric charge and magnetic moment (for example~\cite{Jackson}). A real component of field describes TR of electric charge, while an imaginary part corresponds to magnetic moment. Terms $B_i$, contained in formulae~(\ref{eq:0}) - (\ref{eq:0b}), characterize angular components of current density, induced by magnetic moment of particle. Common multiplier $C_1$ includes a term, describing a spherical wave with frequency $\omega$ at distance $R$ from the source. It should be noted, that polar angle $\theta '$, included in expression for unit wave vector ${\bf n}$, is calculated from negative direction of $z$ axis and is connected with polar angle $\theta$ by a simple relation: $\theta ' = \pi - \theta$.

In order to calculate the circular polarization degree (Stokes parameter $\xi_3$), one needs to use field components defined in a plane perpendicular to photon momentum $\bf k$:

\begin{equation}\label{eq:2}
\xi_3 = i\frac{E_{1}^{*}E_{2} - E_{1}E_{2}^{*}}{|E_1|^2 + |E_2|^2}.
\end{equation}

To accomplish this, the radiation field components $E_1, E_2$ have to be defined in the reference frame with unit polarization vectors ${\bf e}_1, {\bf e}_2$~\cite{Potylitsyn}: 

\begin{align}
   {\bf e}_1 &= \frac{[{\bf n},{\bf b}]}{|[{\bf n},{\bf b}]|} = \frac{\{-n_y; n_x; 0 \}}{\sqrt{1 - n_{z}^{2}}}, \\
   {\bf e}_2 &= [{\bf n},{\bf e}_1] = \frac{\{- n_{x} n_{z};- n_{y} n_{z}; 1 - n_{z}^{2} \}}{\sqrt{1 - n_{z}^{2}}}.
\end{align}

After simple transformations one can get:
\begin{align}
  E_i = \left({\bf E}{\bf e}_i\right) = C_{1} \left(e E_{ei} + i\mu E_{\mu i} \right),
\end{align}

where all values of $E_{e, \mu, i}$ are real:

\begin{align*}
E_{e1} &= - \frac{\beta_{y} \beta_{z} n_{x} n_{z}}{\sqrt{1 - n_{z}^{2}}}, \\
E_{\mu 1} &= \frac{\omega}{c}\frac{\beta_{z} n_{z}^{2} (B_{x} n_{x} + B_{y} n_{y}) - B_{z} (1 - \beta_{y} n_{y}) (1 - n_{z}^{2})}{\gamma \sqrt{1 - n_{z}^{2}}}, \\
E_{e2} &= - \frac{\beta_{z} (1 - \beta_{y} n_{y} - n_{z}^{2})}{\sqrt{1 - n_{z}^{2}}}, \\
E_{\mu 2} &= \frac{\omega}{c}\frac{\beta_{z} n_{z} (B_{y} n_{x} - B_{x} n_{y})}{\gamma \sqrt{1 - n_{z}^{2}}}.
\end{align*}

For the geometry under consideration the denominator in~(\ref{eq:2}) is expressed as follows:

\begin{multline}\label{eq:3}
|E_1|^2 +|E_2|^2 = \frac{|C_1|^2}{(1 - n_{z}^{2})}  \Bigl(e^{2}\beta_{z}^{2}\Bigl[\beta_{y}^{2} n_{x}^{2} n_{z}^{2} + (1 - \beta_{y} n_{y} - n_{z}^{2})^2\Bigr] + \frac{\omega^2}{\gamma^{2} c^2}\mu^2 \cos^{2}\psi \times \\ 
\times \Bigl[n_{x}^{2}\Bigl\{\sin\psi (1 + \gamma)(1 - \beta_{y} n_{y} - n_{z}^{2}) +\gamma^{-1}n_{z}^{2}\sin\psi + \beta n_{z}^{2} n_{y} (1 - \cos\psi)\Bigr\}^2 + \\ 
+ n_{z}^{2}\Bigl\{\gamma \beta_{y} n_{x}^{2}\sin\psi - \beta_{z} n_{x}^{2}\cos\psi - \beta_{z} n_{y}^{2} + \gamma^{-1} n_{y} \sin\psi \Bigr\}^2 \Bigr] \Bigr). 
\end{multline}

As expected, expression~(\ref{eq:3}), describing the TR intensity, is a sum of two terms: the former is proportional to the charge squared and describes the ordinary transition radiation, and the latter, which is proportional to the magnetic moment squared, describes the radiation of the longitudinal magnetic moment. The intensity of the longitudinal magnetic moment radiation does not depend on its orientation (parallel or antiparallel to momentum $\bf p$). For the perpendicular passage, the intensity of magnetic moment radiation was obtained in works~\cite{Ginzburg,Sakuda}. The expression~(\ref{eq:3}) coincides with results from~\cite{Ginzburg,Sakuda} for the geometry $\psi = 0$ $(\beta_z = \beta)$.

Let's consider radiation in the plane of particle incidence ($n_x = 0$) when direction cosines $n_y$, $n_z$ are expressed only through polar angle $\theta '$: $n_y = \sin\theta '$, $n_z = \cos\theta '$. In this case a denominator in formula~(\ref{eq:3}) (see expression for $C_1$) can be written in the following form: 
\begin{multline*}
\Bigl[1 - \beta(\sin\psi\sin\theta ' - \cos\psi\cos\theta ')\Bigr]^2 \Bigl[1 - \beta(\sin\psi\sin\theta ' + \cos\psi\cos\theta ')\Bigr]^2 = \\ 
= \Bigl[1 - \beta\cos(\psi - \theta ')\Bigr]^2 \Bigl[1 + \beta\cos(\psi + \theta ')\Bigr]^2. 
\end{multline*}
As $0 < \theta ' \leq \pi/2$, then maximum values of radiation in the plane of incidence ($\phi = \pi/2$) are determined from the following condition:
\begin{equation*}
   \sin\theta ' = \beta\sin\psi \pm \beta^{-1} \sqrt{(1 - \beta^2)(1 - \beta^2 \sin^2 \psi)}.
\end{equation*}
Angular domains which obey a condition $\sin\theta ' = \beta\sin\psi$, correspond to the minimum in spectral-angular distribution of TR.  For relativistic electrons the spectral-angular distribution of TR has a symmetrical volcano shape with a minimum in a mirror reflection direction and two maxima falling at angles $\theta ' \sim \gamma^{-1}$ relative to mirror reflection direction. With the decrease of electron energy the distribution takes an asymmetrical form and is greatly expanded. TR of electric charge is linearly polarized, but due to the contribution of magnetic moment under oblique incidence in the minimum domain of TR a circular polarized component occurs.
 
The numerator in~(\ref{eq:2}) is defined in the same way: 

\begin{multline}\label{eq:4}
i\left(E_{1}^{*}E_{2} - E_{1}E_{2}^{*}\right) = 2 |C_1|^{2} \frac{\omega}{\gamma c} \frac{e \mu \beta_{y} n_{x} \cos\psi}{(1 - n_{z}^{2})}\times \\
\times \Bigl(\beta_{z} n_{z} \Bigl[\gamma \beta_{y} n_{x}^{2}\sin\psi - \beta_{z} n_{x}^{2}\cos\psi - \beta_{z} n_{y}^{2} + \gamma^{-1} n_{y} \sin\psi \Bigr] + \\
+ (1 - \beta_{y} n_{y} - n_{z}^{2})\Bigl[n_{z}^{2}(\gamma^{-1} \cos\psi + \beta n_{y} \ctg\psi (1 - \cos\psi)) + \\
+ (1 + \gamma)(1 - \beta_{y} n_{y} - n_{z}^{2})\cos\psi \Bigr] \Bigr).
\end{multline}

It is noteworthy that factor $\mu \beta_y n_x$ in numerator~(\ref{eq:4}) just corresponds to mixed product
\begin{equation}
\Bigl({\boldsymbol \mu} [{\bf b},{\bf n}] \Bigr) = \frac{\mu}{\beta}\Bigl({\boldsymbol \beta} [{\bf b},{\bf n}]  \Bigr).
\end{equation}
\begin{figure}[h]
\begin{minipage}[h]{0.49\linewidth}
\center{\includegraphics[width=1\linewidth]{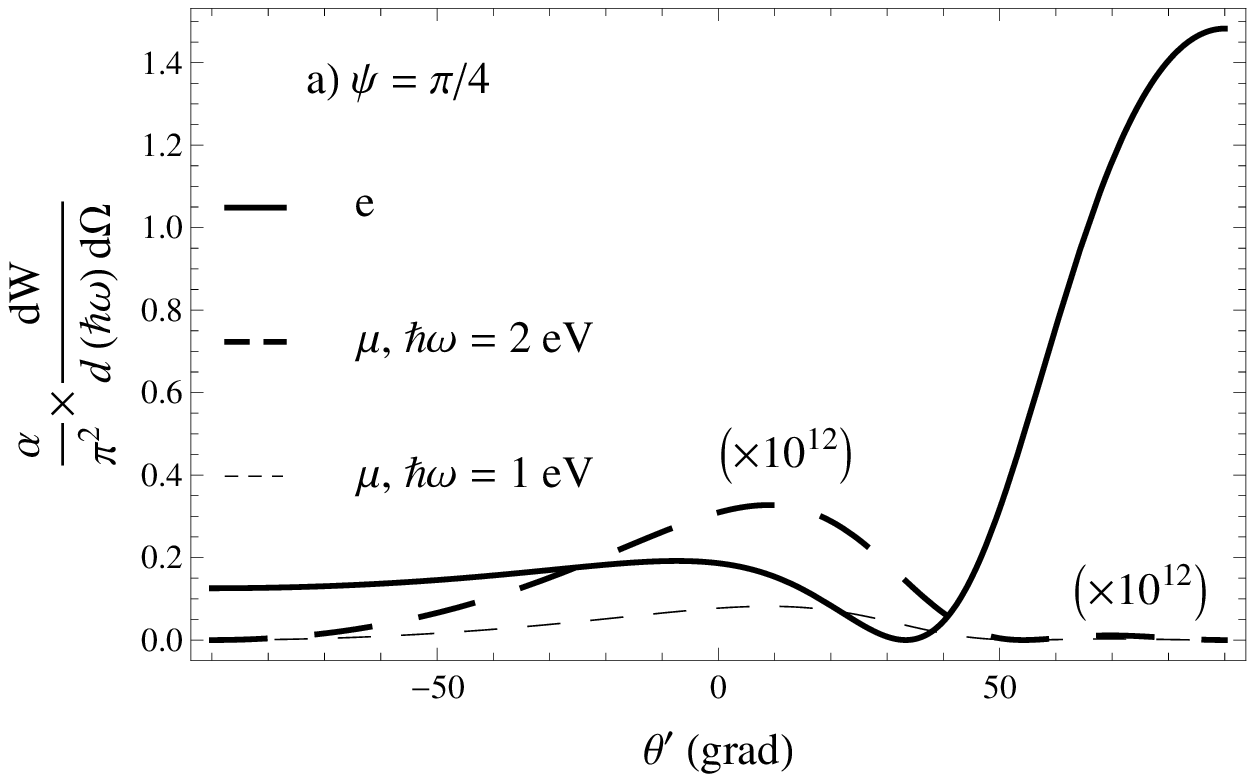}}
\end{minipage}
\hfill
\begin{minipage}[h]{0.49\linewidth}
\center{\includegraphics[width=1\linewidth]{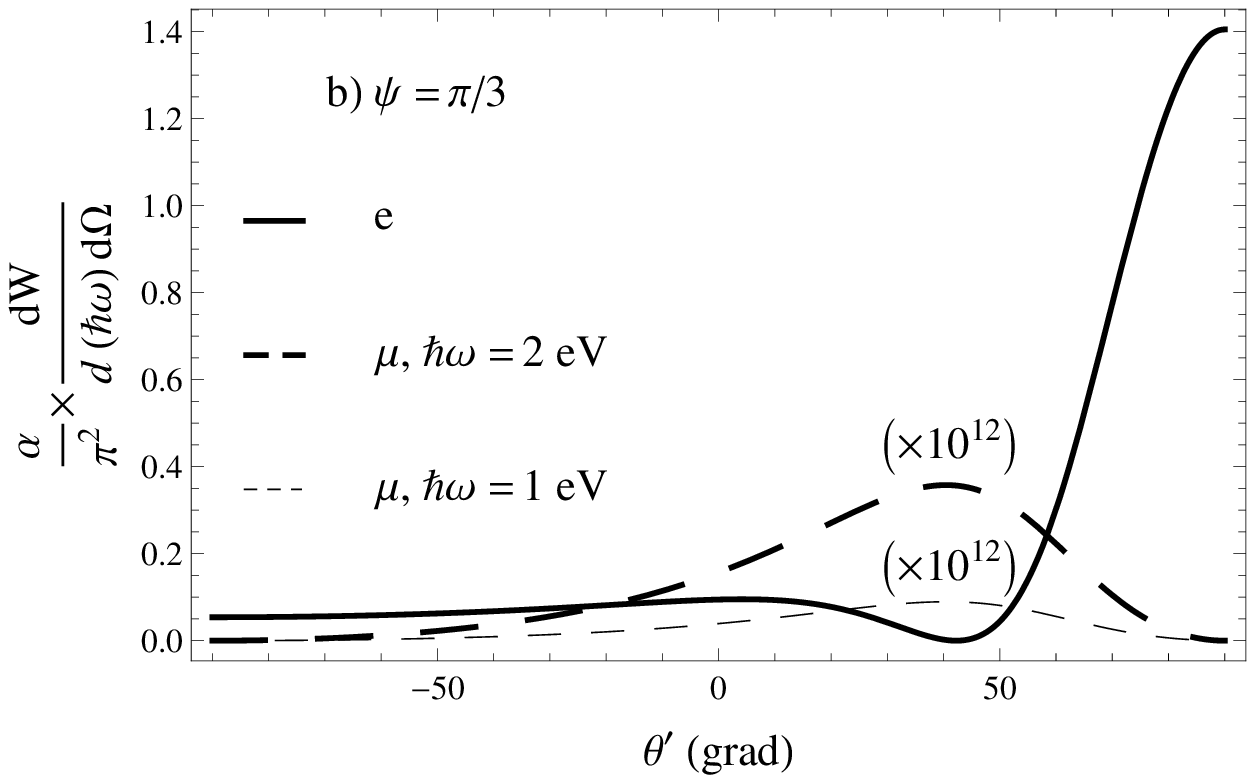}}
\end{minipage}
\caption{Angular TR distribution for electrons with energy 
$E_k = 300$\,keV, with magnetic moment $\mu = \mu_B$, from a perfectly conducting target with $\phi = \pi/2$ (the intensity of the magnetic moment TR is depicted $10^{12}$ times higher for convinince). The solid line represents the ordinary TR defined by a charge; the dashed line, the magnetic moment TR for the photons with energy $\hbar \omega_1 = 2$\,eV; the dotted line, the magnetic moment TR for the photons with energy $\hbar \omega_2 = 1$\,eV}
\label{Fig2}
\end{figure}

Figure~2 shows the TR angular distributions for a perfectly conducting target and incident angles $\psi = \pi/4$ and $\psi = \pi/3$ for kinetic electron energies $E_k = 300$\,keV $(\gamma = 1.587)$ for two photon energies $\hbar \omega_1 = 2$\,eV and $\hbar \omega_2 = 1$\,eV.
The radiation defined by a magnetic moment leads to the situation, when the minimum in the radiation intensity will be non-zero (unlike the ordinary TR). For a perfectly conducting target, the ordinary TR characteristics do not depend on the energy of photons emitted. For the TR from a magnetic moment, there is a dependence on frequency (see the second summand in~(\ref{eq:3})); however the Bohr magneton radiation intensity is $\left(\hbar \omega/\gamma m c^2\right)^2$ times smaller than the charge TR intensity~\cite{Konkov}, so the total intensity virtually does not depend on the frequency.
\begin{figure}[h]
\begin{minipage}[h]{0.49\linewidth}
\center{\includegraphics[width=1\linewidth]{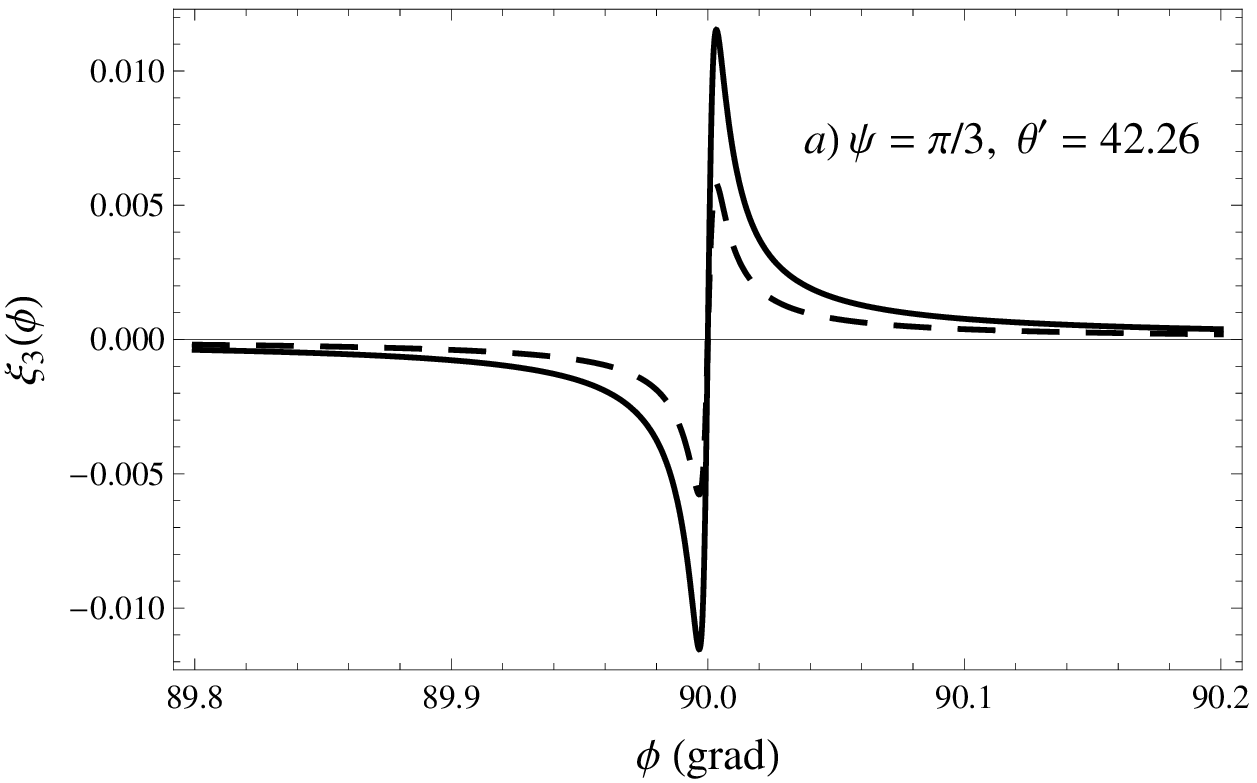}}
\end{minipage}
\hfill
\begin{minipage}[h]{0.49\linewidth}
\center{\includegraphics[width=1\linewidth]{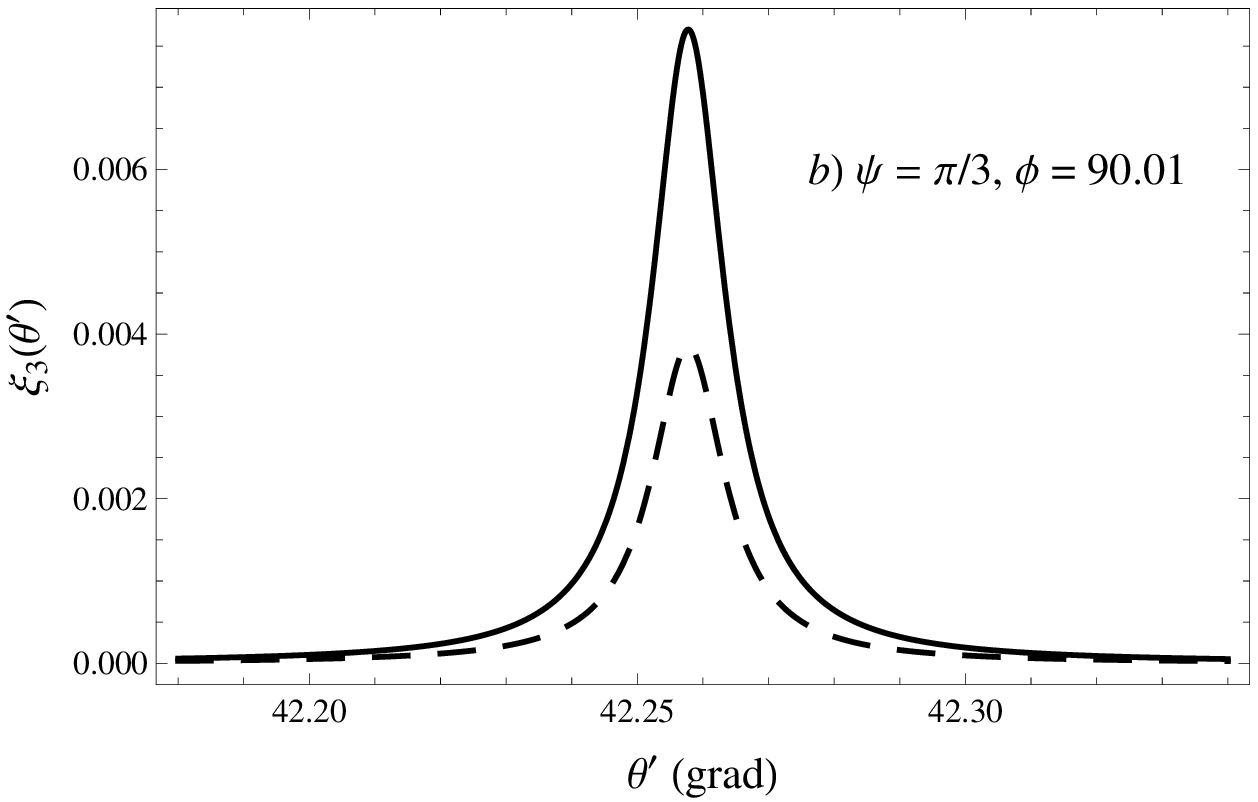}}
\end{minipage}
\caption{Dependence of Stokes parameter $\xi_3$ on polar $\theta '$ and azimuthal $\phi$ angles of TR photons emitted with $\hbar \omega_1 = 2$\,eV (solid line) and $\hbar \omega_2 = 1$\,eV (dashed line) from perfectly conducting target for $\mu = \mu_B$}
\label{Fig3}
\end{figure}

Figure~3 shows the dependence of Stokes parameter $\xi_3$ on polar and azimuthal angles of the TR photons emitted with the same energy. The calculations were performed for the target tilt angle $\psi = \pi/3$, since in this case the valley region of the TR is close to $\theta = 45^\circ$, which leads to the increase of expression~(\ref{eq:1}).
\begin{figure}[htb]
\begin{minipage}[htb]{0.49\linewidth}
\center{\includegraphics[width=1\linewidth]{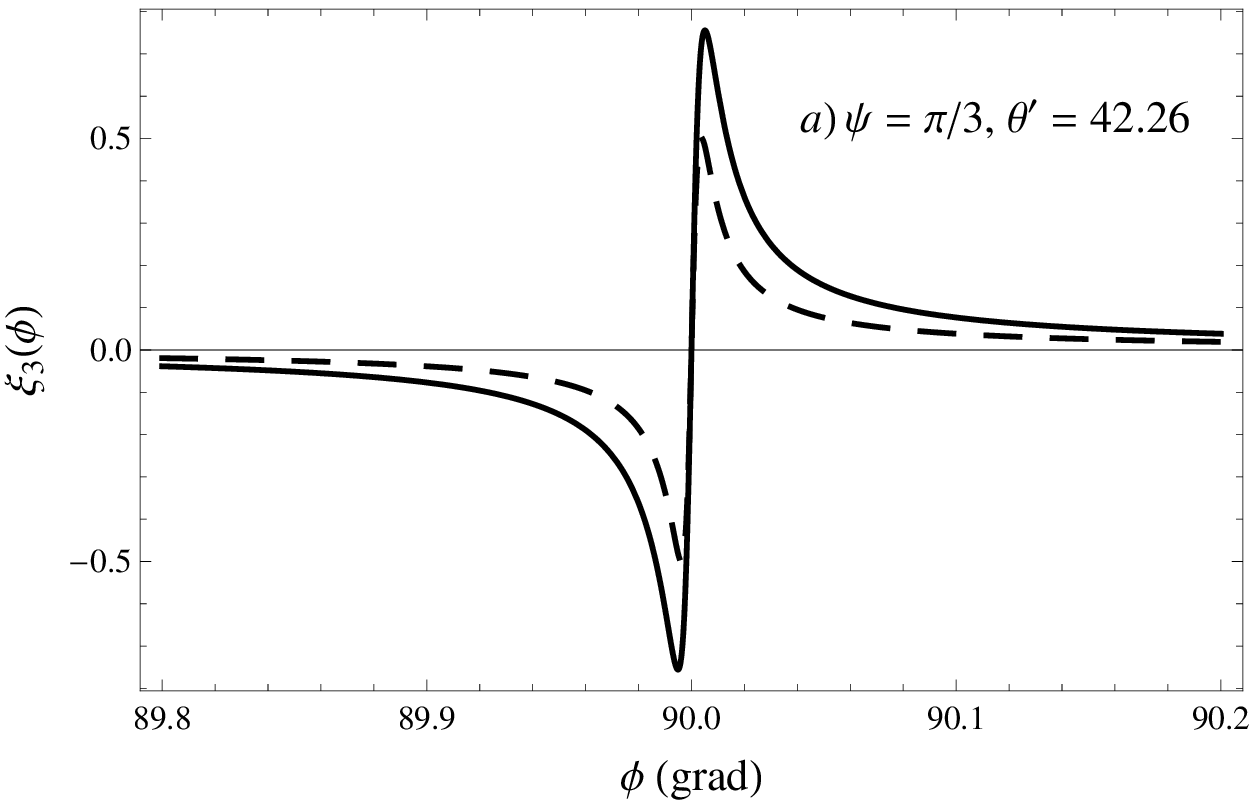}}
\end{minipage}
\hfill
\begin{minipage}[htb]{0.49\linewidth}
\center{\includegraphics[width=1\linewidth]{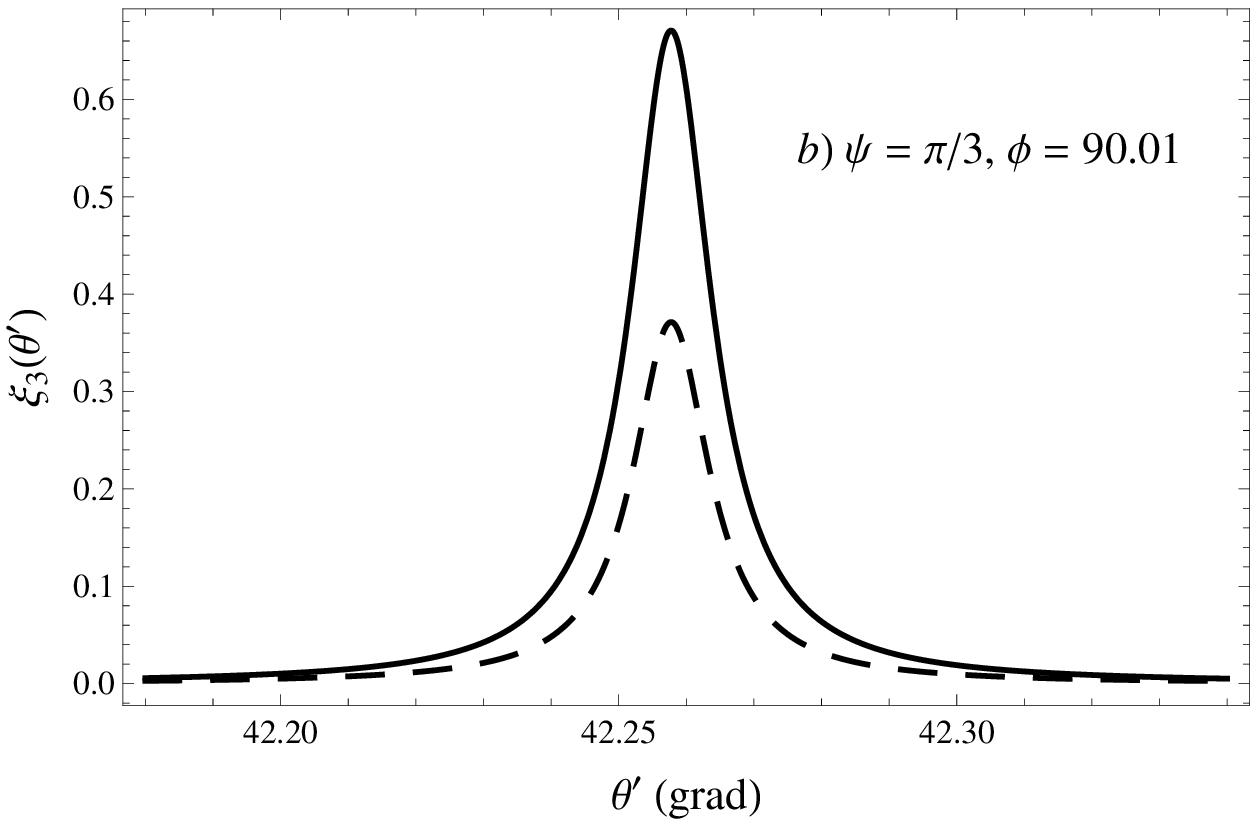}}
\end{minipage}
\caption{Dependence of Stokes parameter $\xi_3$ on polar $\theta '$ and azimuthal $\phi$ angles of TR photons emitted with $\hbar \omega_1 = 2$\,eV (solid line) and $\hbar \omega_2 = 1$\,eV (dashed line) from perfectly conducting target for $\mu = 100 \mu_B$}
\label{Fig4}
\end{figure}

As one can see in the figure, circular polarization does not exceed 1\% for the TR of a longitudinally spin-polarized electron. However, this value increases significantly for electrons with OAM ($\ell \gg \hbar$) (see figure~4).

It should be mentioned that in the region of minimum radiation intensity (near $\theta ' = 42.26^\circ$, see figure~2b) the circular polarization of radiation for electrons with $\ell \approx 100\hbar$ can reach the values of 70\% depending on radiation angles (see figure~4).

\section{Discussion and conclusions}
Hence, one can measure the OAM value by measuring the circular polarization of optical transition radiation from electrons with OAM in the range of the angles that correspond to the minimum TR intensity, and the accuracy of such optical measurements can be better that 1\%. 

It should be noted that this value of polarization degree is observed in a very narrow angular range of the order of tenths - hundredth of a degree in the region of minimum of spectral-angular distribution of TR intensity. Hence direct detection of this effect is impossible on the modern level of technology development of vortex electron beams creation. Therefore the most advanced method to measure a degree of circular polarization of TR in the noted range is the use of optical systems. For example, in work~\cite{Kalinin} the authors proposed to use a parabolic telescope to exclude the effect of pre-wave zone. The parabolic telescope is a system which is basically a parabolic mirror with a detector placed into the focus of a mirror. The using of this approach will allow measuring a degree of circular polarization of TR even for very small ranges of radiation angles. Thus, for example, for a beam with diameter of 1 cm ($10^4 \lambda$) and parabolic mirror with radius 2 cm (focus distance is 1 cm) the displacement of a beam for $0.2^\circ$ leads to the image formation on the focal surface with size of $\Delta = 350\,\mu m$. In this case value $\Delta$ greatly exceeds the value of diffraction limit $\delta\lambda = 0.6 - 1.6\, \mu m$. (for energy of photon $\hbar \omega = 1 - 2$ eV). Using a slot with aperture of $200\, \mu m$ one can precisely measure the degree of circular polarization in the region of minimum of spectral-angular distribution of TR, neglecting the influence of maxima in distribution.   

The TR characteristics for polished aluminum in the optical region are close to the calculated ones for perfect conductor~\cite{Wartski, bib:P}. The evaluations obtained hereinabove are expected to be valid for a polished metal target.  

The authors of work~\cite{Sakuda:Kurihara} have shown that the strict consideration of the quantum corrections leads to significant increase of Bohr magneton TR intensity and which was calculated using classical electrodynamics. But if magnetic momentum $\sim$ 100-fold exceeds the Bohr magneton, as in the case under consideration, one should expect that the quantum corrections will be insignificant.

\section{Acknowledgements}
Authors are thankful to V.\,G. Serbo, I.\,P. Ivanov and D.\,V. Karlovets for inspiring discussions. The work was partially supported by Russian Ministry of Science and Education within the grants No. 14.B37.21.0775, No. P1199 and Program ``Nauka''.


\begin{thebibliography}{18}

\bibitem{Nye}
J.\,F. Nye, M.\,V. Berry, Proc. R. Soc. Lond. A, 336 (1974), p. 165.

\bibitem{Allen}
L. Allen, M.\,W. Beijersbergen, R.\,J.\,C. Spreeuw and J.\,P. Woerdman, Phys. Rev. A, 45 (1992), p. 8185.

\bibitem{bib:B}
K.\,Y. Bliokh, Y.\,P. Bliokh, S. Savel'ev and F. Nori, Phys. Rev. Lett., 99 (2007), p. 190404.

\bibitem{Uchida}
M. Uchida, A. Tonomura, Nature, 464 (2010), p. 737.

\bibitem{Verbeeck}
J. Verbeeck, H. Tian, P. Schattschneider, Nature, 467 (2010), p. 301.

\bibitem{McMorran}
B.\,J. McMorran et al., Science, 331 (2011), p. 192.

\bibitem{Potylitsyn}
A.\,P. Potylitsyn, Electromagnetic Radiation of Electrons in Periodic Structures, Springer, 2011.

\bibitem{Bliokh}
K.\,Y. Bliokh, M.\,R. Dennis and F. Nori, Phys. Rev. Lett., 107 (2011), p. 174802.

\bibitem{Bal}
C. Bal et al., Proceeding DIPAC 2003 - Mainz, Germany, PM04, p. 95. 

\bibitem{Ginzburg}
V.\,L. Ginszburg, V.\,N. Tsytovich, Transition Radiation and Transition Scattering, A. Hilger, Bristol, 1990.

\bibitem{Konkov}
A.\,S. Konkov, A.\,P. Potylitsyn, V.\,A. Serdutskiy, Russian Physics Journal, 54 (2012), p. 1249.

\bibitem{Pafomov}
V.\,E. Pafomov, Proc. FIAN USSR, 44 (1969), p. 90.

\bibitem{Jackson}
J.\,D. Jackson, Classical Electrodynamics, Wiley, New
York, 1962.

\bibitem{Sakuda}
M. Sakuda, Phys. Rev. Lett., 72 (1994), p. 804.

\bibitem{Wartski}
L. Wartski, S. Roland, J. Lassale et al., J. Appl. Phys., 46 (1975), p. 3644.

\bibitem{bib:P}
A.\,P. Potylitsyn, Nucl. Instrum. and Meth. B, 145 (1998), p. 169.

\bibitem{Sakuda:Kurihara}
M. Sakuda and Y. Kurihara, Phys. Rev. Lett., 74 (1995), p. 1284.

\bibitem{Kalinin}
B.\,N. Kalinin, G.\,A. Naumenko, A.\,P. Potylitsyn et al., JETP Lett., 84 (2006), p. 110.
\end{thebibliography}
\end{document}